\documentclass[letterpaper, 10 pt, conference]{ieeeconf}  

\IEEEoverridecommandlockouts                              

\usepackage{graphics} 
\usepackage{amsmath} 
\usepackage{amssymb}  
\usepackage{graphicx}
\usepackage{dsfont}
\usepackage{booktabs}
\usepackage{bm}
\usepackage{makecell} 
\usepackage{cite}

\title{\LARGE \bf
EEG-Based Inter-Patient Epileptic Seizure Detection Combining Domain Adversarial Training with CNN-BiLSTM Network
}

\author{Rina Tazaki$^{1}$, Tomoyuki Akiyama$^{2}$, and Akira Furui$^{1}$
\thanks{This work was partially supported by JSPS KAKENHI Grant Number JP23K28128.}
\thanks{$^{1}$R. Tazaki and A. Furui are with the Graduate School of Advanced Science and Engineering, Hiroshima University, Higashi-hiroshima, Japan
        (e-mail: \{rinatazaki, akirafurui\}@hiroshima-u.ac.jp).}%
\thanks{$^{2}$T. Akiyama is with the Department of Pediatric Neurology, Okayama University Hospital, Okayama, Japan (e-mail:  takiyama@okayama-u.ac.jp).}%
}

\begin{document}

\maketitle
\thispagestyle{empty}
\pagestyle{empty}

\begin{abstract}
Automated epileptic seizure detection from electroencephalogram (EEG) remains challenging due to significant individual differences in EEG patterns across patients.
While existing studies achieve high accuracy with patient-specific approaches, they face difficulties in generalizing to new patients.
To address this, we propose a detection framework combining domain adversarial training with a convolutional neural network (CNN) and a bidirectional long short-term memory (BiLSTM). 
First, the CNN extracts local patient-invariant features through domain adversarial training, which optimizes seizure detection accuracy while minimizing patient-specific characteristics.
Then, the BiLSTM captures temporal dependencies in the extracted features to model seizure evolution patterns.
Evaluation using EEG recordings from 20 patients with focal epilepsy demonstrated superior performance over non-adversarial methods, achieving high detection accuracy across different patients. 
The integration of adversarial training with temporal modeling enables robust cross-patient seizure detection. 

\end{abstract}

\section{Introduction}

Epileptic seizures are neurological symptoms caused by abnormal electrical activity in cerebral neurons.
The standard method for detecting these seizures  involves analyzing electroencephalogram (EEG) recordings obtained from the scalp surface.
However, since seizures may not occur during typical EEG sessions, which usually last several tens of minutes, multiple recordings are often necessary~\cite{faulkner2012utility}.
In some cases, long-term video-EEG monitoring may be required.
The manual review of these extensive recordings by epileptologists is extremely time-consuming and labor-intensive.

To address this clinical burden, various automated seizure detection approaches based on machine learning have been developed.
Many of these methods adopt patient-specific classification approaches, where models are trained with EEG data from individual patients~\cite{cite:shoeb_patient-specific_2004, cite:kaleem_patient-specific_2018, cite:hunyadi2012}.
While these approaches achieve high detection accuracy by optimizing models for individuals, they require collecting sufficient seizure data and training new models for each patient, limiting their clinical applicability.

Inter-patient seizure detection approaches have emerged to develop more generalized models by learning from EEG data collected from multiple patients~\cite{cite:craley_automated_2021, cite:thodoroff2016, alalayah2023effective}.
However, many existing studies have not explicitly addressed individual differences in EEG patterns, leading to decreased detection accuracy when the EEG features of new patients differ significantly from the training data.
This phenomenon, known as domain shift, necessitates treating each patient as a distinct domain and accounting for inter-domain distributional differences to achieve robust seizure detection.

In recent years, domain generalization through adversarial training has emerged as a promising approach to address the domain shift problem in inter-patient seizure detection. 
However, existing studies have primarily relied on convolutional neural network (CNN)-based architectures~\cite{cite:zhang_adversarial_2020, peng2022domain, zhang2024cross}, which are limited to capturing local patterns within the time-series EEG data.
While these approaches effectively learn patient-invariant features, they do not adequately capture the temporal evolution of seizure patterns, which is crucial for accurate detection across different patients.

This paper proposes an epileptic seizure detection method that integrates the concept of domain-adversarial neural networks~\cite{cite:ganin2015unsupervised} with temporal modeling to address individual differences in EEG patterns.
The proposed method combines a CNN to extract local spatiotemporal features and a bidirectional long short-term memory (BiLSTM)~\cite{graves2005framewise} to capture long-term temporal dependencies. 
Through adversarial training, the model learns features that are both patient-invariant and temporally informative, enabling robust seizure detection that generalizes across individual patient variations.

\section{Proposed Method}

\begin{figure*}[htb] 
    \centering
    \includegraphics[width=0.7\linewidth]{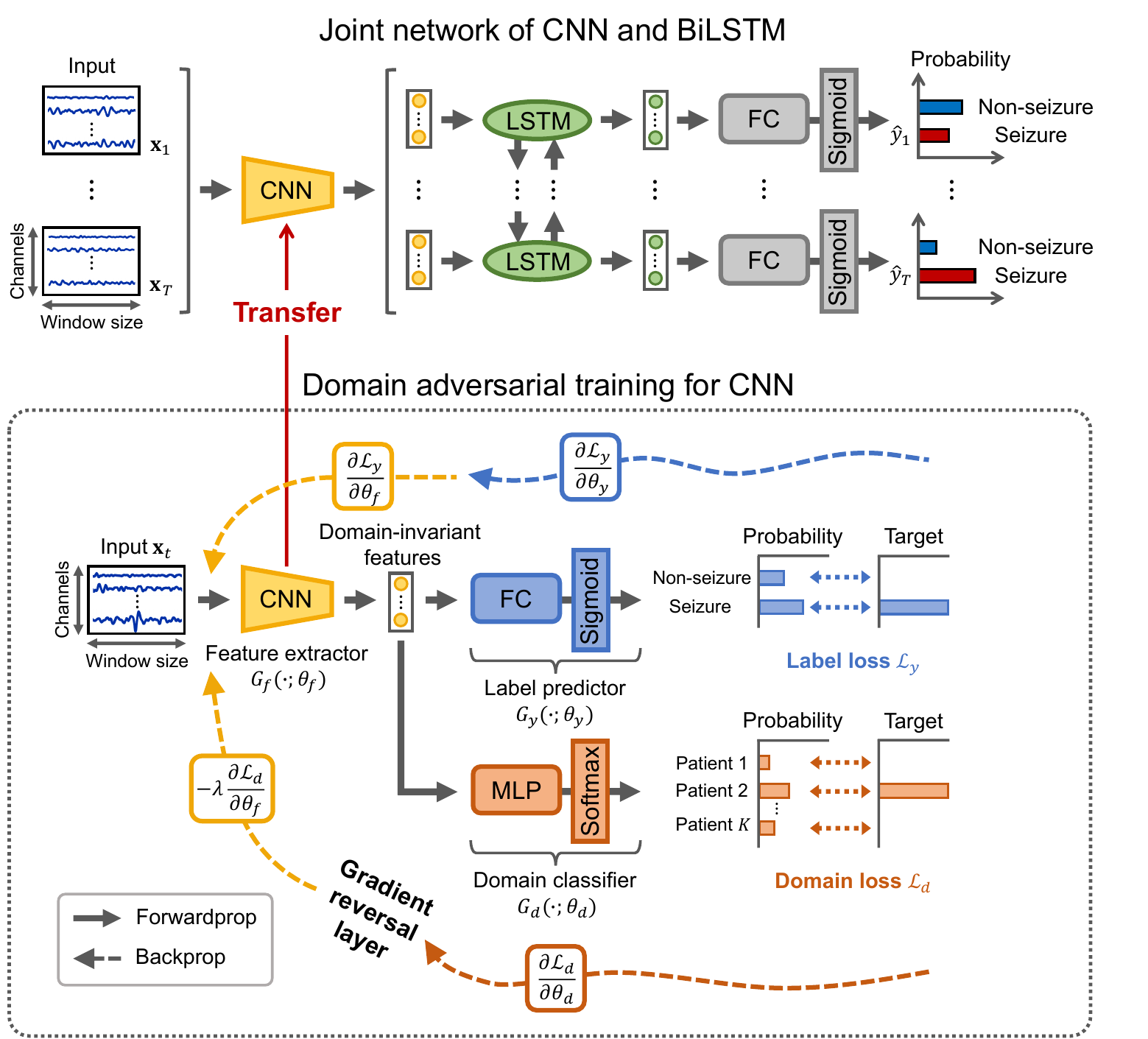} 
    \caption{Schematic overview of the proposed CNN-BiLSTM framework with adversarial training} 
    \label{fig:method}
\end{figure*}

\subsection{Problem Setting}

We assume that long-term EEG recordings are pre-processed and segmented into windows of length $L$.
Under this assumption, we consider the EEG signal $\mathbf{x}_t \in \mathbb{R}^{C \times L}$ at time $t$ (where $C$ denotes the number of channels and $L$ represents the window length) and its corresponding class label $y_t \in \{0, 1\}$ (0: non-seizure, 1: seizure).
For inter-patient epileptic seizure detection, data from multiple source patients are assumed to be available for model training.
Given that $T^k$ represents the length of the time-series for patient $k$, the dataset for the $k$-th patient can be expressed as $\mathcal{D}^k = \{(\mathbf{x}_t^k, y_t^k)\}_{t=1}^{T^k}$.
With $K$ source patients, the entire source dataset is denoted as $\mathcal{D} = \{\mathcal{D}^k\}_{k=1}^K$.
The model trained on $\mathcal{D}$ is subsequently used to detect seizures in the EEG time-series $\{\mathbf{x}_t^\mathrm{trg}\}$ of a target patient.

\subsection{Network Architecture}

Fig.~\ref{fig:method} shows the overview of the proposed method. 
Our approach employs a CNN to extract discriminative features from EEG recordings at specific time points, followed by a BiLSTM to capture the temporal information of these features, enabling seizure detection from continuous EEG signals.
However, the CNN-BiLSTM architecture alone cannot explicity account for inter-patient differences in data distribution, potentially leading to decreased accuracy due to individual differences in EEG patterns.
To address this challenge, we introduce a two-stage learning process that incorporates adversarial training.

In the first stage, domain adversarial training~\cite{cite:ganin2015unsupervised} is applied to the CNN, which serves as a feature extractor. 
Specifically, a domain classifier is introduced in addition to the feature extractor and a label predictor, training the feature extractor to adversarially deceive the domain classifier.
This approach enables the feature extractor to learn domain-invariant (i.e., patient-invariant) features, potentially reducing the impact of individual differences.
In the second stage, we fine-tune the adversarially trained CNN together with the BiLSTM.
This process allows the model to extract patient-invariant features while considering time-series information, thereby improving its generalization capability.
Consequently, the proposed method enables robust seizure detection across different patients.

\subsection{Domain Adversarial Training for CNN}
In the first stage, adversarial training is applied to the CNN to capture patient-invariant features. 
The entire source dataset $\{\mathcal{D}^k\}_{k=1}^K$ can be reformulated as $\mathcal{D} = \{(\mathbf{x}_i, y_i, d_i)\}_{i=1}^N$ by combining data from all source patients and incorporating domain information for each data point.
Here, $d_i \in \{1, \ldots, K\}$ represents the domain (patient) label, and $N = \sum_{k=1}^K T^k$ denotes the total number of data points across all source patients.
During training, we generate mini-batches of size $N_b$ by sampling from different patients and time points within this dataset.

The first stage of training involves connecting a label predictor $G_y$ and domain classifier $G_d$ to the CNN-based feature extractor $G_f$. 
We denote the parameters of $G_f$, $G_y$, and $G_d$ as $\theta_f$, $\theta_y$, and $\theta_d$, respectively. 
Since each domain corresponds to an individual patient, the adversarial training between the feature extractor and domain classifier facilitates the extraction of patient-invariant features, thereby improving generalization to the target patient.
In the proposed method, the label predictor $G_y$ consists of a fully connected (FC) layer with a sigmoid activation function, while the domain classifier $G_d$ is implemented as a multilayer perceptron (MLP) with a softmax activation function.

The binary cross-entropy loss $\mathcal{L}_y$ for label prediction is formulated as follows:
\begin{equation}
    \mathcal{L}_y(\theta_f, \theta_y) = -\frac{1}{N_b} \sum_{i \in N_b} w_{y_i} \left[y_i \ln \hat{y}_i + (1 - y_i) \ln (1 - \hat{y}_i)\right],
\label{Loss_y}
\end{equation}
where $\hat{y}_i = G_y(G_f(\mathbf{x}_i; \theta_f); \theta_y)$ represents the seizure probability predicted by $G_y$ based on features extracted by $G_f$. 
The term $w_{y_i}$ denotes class-specific weighting coefficients.
To address the inherent class imbalance between seizure and non-seizure classes, we define $w_{y_i}$ as:
\begin{align}
    w_0 &= \frac{N}{2 \sum_{i=1}^N (1 - y_i)}, \label{w_0} \\
    w_1 &= \frac{N}{2 \sum_{i=1}^N y_i}. \label{w_1}
\end{align}
The loss function $\mathcal{L}_y$ decreases as the label predictor $G_y$ assigns higher probability to the true label $y_i$.
For domain classification, we define the cross-entropy loss $\mathcal{L}_d$ as:
\begin{equation}
	\mathcal{L}_d(\theta_f, \theta_d) = -\frac{1}{N_b} \sum_{i \in N_b} \sum_{k=1}^K \mathds{1}_{[k=d_i]} \ln G_d^{(k)}(G_f(\mathbf{x}_i; \theta_f); \theta_d).
\label{Loss_d}
\end{equation}
Here, $\mathds{1}_{[k=d_i]}$ denotes an indicator function that equals 1 when $k=d_i$ and 0 otherwise.
Consequently, only the predicted probability for the true domain contributes to the loss computation.

The overall optimization objective is formulated as follows:
\begin{equation}
\min_{\theta_f, \theta_y} \max_{\theta_d} \mathcal{L}_y(\theta_f, \theta_y) - \lambda \mathcal{L}_d(\theta_f, \theta_d),
\label{minmax}
\end{equation}
where $\lambda$ is a hyperparameter controling the contribution of the domain classification loss $\mathcal{L}_d$.
The sign reversal of $\mathcal{L}_d$ is achieved using a gradient reversal layer (GRL) between $G_f$ and $G_d$.
This optimization framework encourages the feature extractor $G_f$ to generate features that reduce the domain classifier's accuracy, while simultaneously training $G_d$ to improve its domain classification performance.
Through this adversarial training process, $G_f$ learns to extract domain-invariant (i.e., patient-invariant) features that are  specifically optimized for seizure detection.

\subsection{End-to-end Training for Entire CNN-BiLSTM}
In the second stage, we integrate a BiLSTM with the domain-adversarially trained CNN to capture temporal dependencies in patient-specific data.
The outputs of the BiLSTM are passed through an FC layer with a sigmoid activation function to generate seizure prediction probabilities.
For this stage, we reconstruct the source dataset as $\mathcal{D} = \{(\mathbf{X}^k, \mathbf{Y}^k)\}_{k=1}^K$, treating the time-series data of each patient as a single sample.
Here, $\mathbf{X}^k = \{\mathbf{x}_t^k\}_{t=1}^{T^k}$ denotes the entire EEG time-series for patient $k$, and $\mathbf{Y}^k=\{y_t^k\}_{t=1}^{T^k}$ represents the corresponding sequence of class labels.
During training, we construct mini-batches of size $N_b'$ by randomly sampling complete patient time-series from this dataset.

For the loss function, we use the following binary cross-entropy loss:
\begin{align}
	\mathcal{L} = -\frac{1}{\sum_{k \in N_b'} T^k} & \sum_{k \in N_b'} \sum_{t=1}^{T^k} w_{y_t^k} \\ \nonumber 
    & \times \left[y_t^k \ln \hat{y}_t^k + (1 - y_t^k) \ln (1 - \hat{y}_t^k)\right],
\label{Loss}
\end{align}
where $\hat{y}_t^k = p(y_t^k = 1|\mathbf{x}_t^k)$ represents the predicted probability of seizure occurrence.
The class-specific weights $w_{y_t^k}$ are calculated following the same methodology as in the domain adversarial training stage to address class imbalance.
The optimization process updates the parameters of all network components: the CNN, BiLSTM, and FC layer.

\subsection{Inference Algorithm}
For inference, we use the trained CNN-BiLSTM model to predict class labels $\{y_t^{\mathrm{trg}}\}$ for the target patient's EEG time-series $\{\mathbf{x}_t^\mathrm{trg}\}$.
The seizure detection at each time point $t$ follows the decesion rule:
\begin{equation}
  \hat{y}_t^\mathrm{trg} =
  \begin{cases}
    1\ (\text{seizure}) & \text{if $p(y_t^\mathrm{trg} = 1 | \mathbf{x}_t^\mathrm{trg}) \geq \tau$,} \\
    0\ (\text{non-seizure}) & \text{otherwise,}
  \end{cases}
\end{equation}
where $p(y_t^\mathrm{trg} = 1 | \mathbf{x}_t^\mathrm{trg})$ denotes the predicted seizure probability at time $t$, and $\tau$ represents a predefined decision threshold ($0 < \tau < 1$).
This binary classification scheme assigns a seizure state when the predicted seizure probability exceeds threshold $\tau$.

\section{Experiments}

\subsection{Dataset and Preprocessing}
In this study, we used a dataset~\cite{furui2020non} comprising EEG recordings from 20 patients diagnosed with focal epilepsy (age range: 0.5--41 years).
The EEG signals were recorded using 21 electrodes positioned according to the international 10--20 electrode placement system, with a sampling frequency of $f_{\mathrm{s}} = 500$ Hz.
We applied longitudinal bipolar montage to all electrodes except the earlobe electrodes (A1, A2), resulting in 18-channel waveforms ($C=18$).
Each patient contributed a single EEG recording containing one seizure event.
The recordings had a mean duration of $319.5 \pm 44.5$ s, with seizures lasting $47.4 \pm 22.7$ s on average.
An expert epileptologist annotated the onset and offset times of each seizure at the individual data point level.
All EEG measurements were conducted with approval from the Okayama University Ethics Committee (approval No: 1706-019). 

To address the common presence of artifacts in raw EEG signals, specifically low-frequency movement artifacts and high-frequency muscle activity, we implemented a bandpass filter (8--30 Hz) as a preprocessing step.
This frequency range was selected to attenuate artifacts while preserving the frequency bands: $\alpha$ (8--12 Hz), $\beta$ (13--24 Hz), and low $\gamma$ (25--30 Hz), which have been demonstrated to be effective for epileptic seizure detection in previous studies~\cite{cite:craley_automated_2021, cite:furui2021time}.
After filtering, we performed channel-wise standardization to achieve zero mean and unit standard deviation for each recording.
The preprocessed signals were then segmented into non-overlapping one-second windows ($L = 500$).
For window-level labeling, we assigned the most frequently occurring class label among the 500 data points in each window as the representative class label.

\subsection{Experimental Conditions}
\subsubsection{Network structure}

The CNN model in the proposed method consists of four blocks. 
Each block contains two identical substructures, comprising a one-dimensional convolutional layer, batch normalization layer, and Leaky ReLU activation function (negative slope of 0.1).
A max pooling layer follows each block to reduce the temporal dimension of feature maps by half. 
The output from the final block passes through a global average pooling layer to generate a one-dimensional feature vector.
For the convolutional layers, we set the kernel size to 3, stride to 1, and padding to 1.
The max pooling layers used a kernel size and stride of 2. 
The number of output channels progressively increased through the blocks (5, 10, 20, and 40), resulting in a 40-dimensional feature vector.

The label predictor transforms the CNN-extracted features into binary predictions (seizure/non-seizure) using two fully connected layers without activation functions, with an intermediate layer size of 20.
The domain classifier employs a three-layer MLP: the first two layers maintain the feature vector dimensionality and include batch normalization and Leaky ReLU activation (negative slope of 0.1).
The final layer outputs $K$ dimensions, where each dimension represents one patient domain.

For temporal modeling of the extracted features, the BiLSTM component incorporates two hidden layers, with 20 units in each layer, to process the windowed data.
The temporal features from BiLSTM are then passed through a single FC layer for final label prediction. 

\subsubsection{Training settings}
In the first stage, we jointly trained the CNN feature extractor, label predictor, and domain classifier using the Adam optimizer with a learning rate of 0.005 and batch size of $N_b = 32$ over 20 epochs.
The hyperparameter $\lambda$ in (\ref{minmax}) was dynamically adjusted according to: 
\begin{equation}
	\lambda = \frac{2}{1 + \exp(-10 p)} - 1,
\label{lambda}
\end{equation}
where $p$ represents the progress of training and gradually changes from 0 to 1.
This value is calculated by dividing the current training step by the product of the total number of epochs and the number of batches per epoch.
This formulation allows $\lambda$ to gradually increase during training, prioritizing seizure detection in early stages while progressively emphasizing domain-invariant feature learning.

In the second stage, we fine-tuned the complete model (CNN, BiLSTM, and FC layer) using Adam optimizer with a learning rate of 0.001 and batch size of $N'_b = 2$ over 6 epochs.
The CNN weights from the first stage were used to initialize the corresponding layers.

\begin{figure}[t] 
    \centering 
    \includegraphics[width=1.0\linewidth]{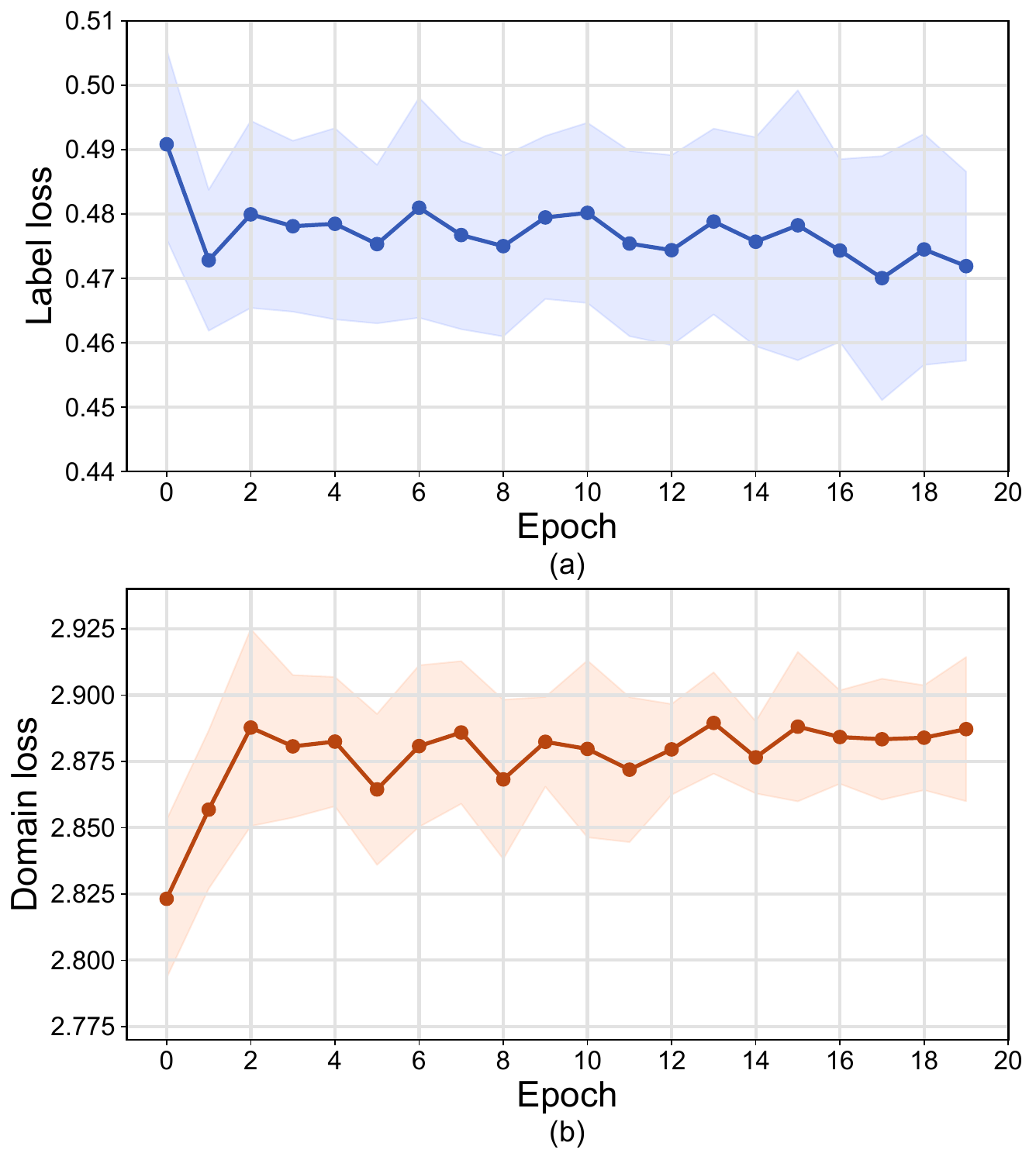} 
    \caption{
        Averaged loss progression during the first-stage training across patients as a function of training epochs. 
        (a) Label loss for seizure prediction. (b) Domain loss for patient classification.
        } 
    \label{fig:Loss} 
\end{figure}

\subsection{Evaluation Methodology}
We employed leave-one-patient-out cross-validation to evaluate the model performance.
In this evaluation scheme, data from 19 patients ($K=19$) served as the source training dataset, while data from the remaining patient were used for testing.
This process was repeated 20 times to cover all patients as the test subject.
The threshold $\tau$ for seizure prediction probabilities was determined by selecting the value that maximized the F1 score on the precision-recall curve using the training data, and this threshold was subsequently applied to the test data.

For performance assessment, we used multiple evaluation metrics: sensitivity, specificity, Matthews correlation coefficient (MCC), area under the receiver operating characteristic curve (AUC-ROC), and area under the precision-recall curve (AUC-PR). 
To ensure robust evaluation, we conducted 15 independent training and evaluation runs with different random seeds and computed the mean value for each metric.

\begin{table*}[tb]
    \centering
    \caption{Results on the effectiveness of adversarial training (AT): mean $\pm$ SD across patients}
    \label{AT_results}
    \begin{tabular}{l|ccccc}
        \toprule
        Method & Sensitivity & Specificity & MCC & AUC-ROC & AUC-PR \\
        \midrule
        CNN \\
        \quad w/o AT & $0.5690 \pm 0.2013$ & $0.9089 \pm 0.0769$ & $0.4629 \pm 0.2073$ & $0.8142 \pm 0.1339$ & $0.5717 \pm 0.2378$ \\
        \quad w/ AT & $0.5703 \pm 0.2274$ & $0.9191 \pm 0.0714$ & $0.4812 \pm 0.2453$ & $0.8035 \pm 0.1369$ & $0.5651 \pm 0.2624$ \\
        CNN-BiLSTM \\
        \quad w/o AT & $0.6890 \pm 0.2603$ & $0.8830 \pm 0.1616$ & $0.5695 \pm 0.2333$ & $0.9205 \pm 0.1238$ & \scalebox{0.92}[1.0]{$\mathbf{0.8020 \pm 0.2002}$} \\
        \quad w/ AT (\textbf{ours}) & \scalebox{0.92}[1.0]{$\mathbf{0.7197 \pm 0.2785}$} & \scalebox{0.92}[1.0]{$\mathbf{0.9297 \pm 0.0747}$} & \scalebox{0.92}[1.0]{$\mathbf{0.6136 \pm 0.2567}$} & \scalebox{0.92}[1.0]{$\mathbf{0.9227 \pm 0.1235}$} & $0.7565 \pm 0.2504$ \\
        \bottomrule
    \end{tabular}
\end{table*}

\begin{figure*}[tb] 
    \centering 
    \includegraphics[width=0.9\linewidth]{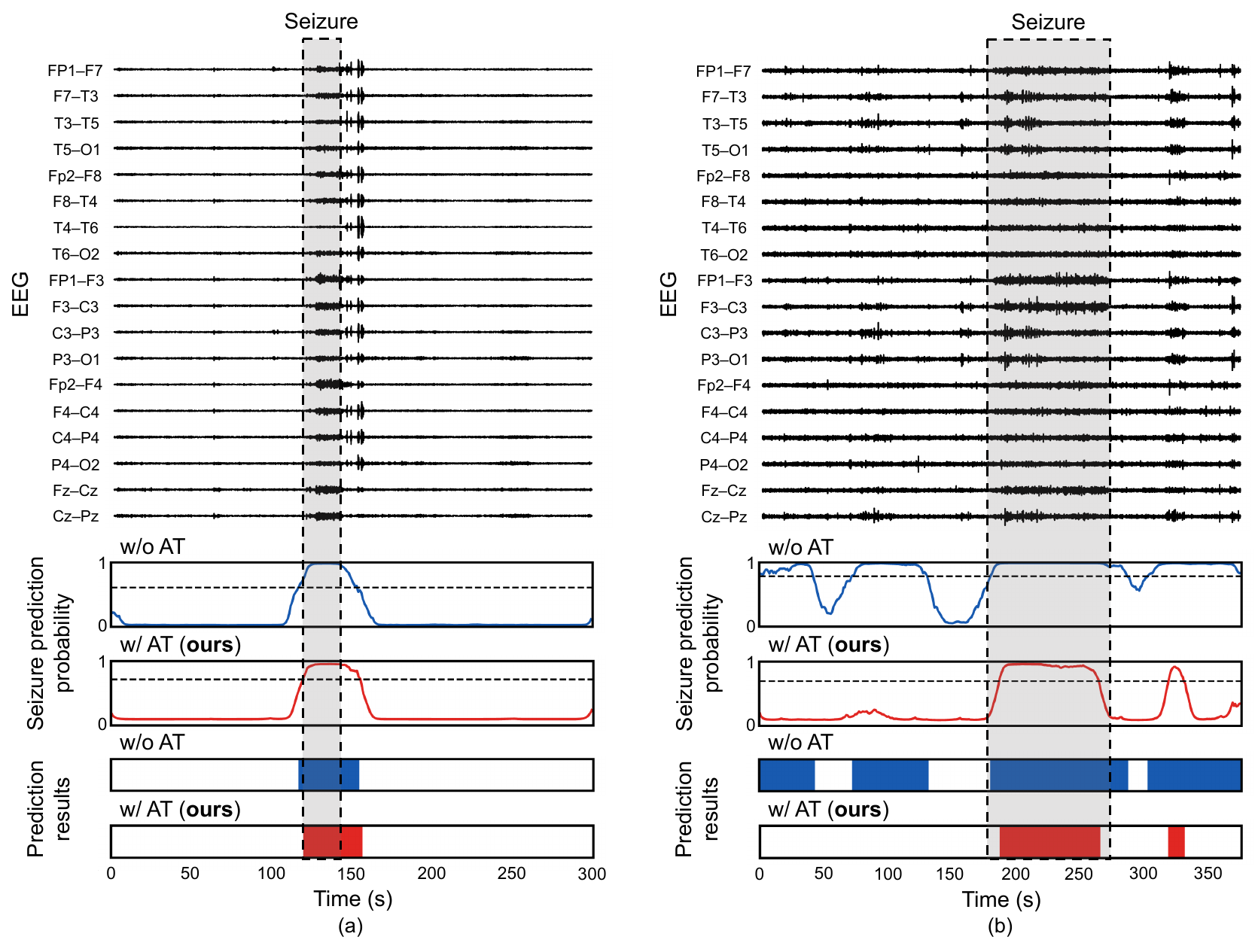} 
    \caption{
    Preprocessed EEG signals and seizure detection results from CNN-BiLSTM models with adversarial training (w/ AT) and without adversarial training (w/o AT).
    (a) Patient K. (b) Patient D.
    Gray regions indicate seizure periods annotated by an epileptologist.
    Horizontal dashed lines in the seizure prediction probability panels represent the detection threshold.
    }
    \label{fig:results}
\end{figure*}

\begin{figure}[t] 
    \centering 
    \includegraphics[width=1.0\linewidth]{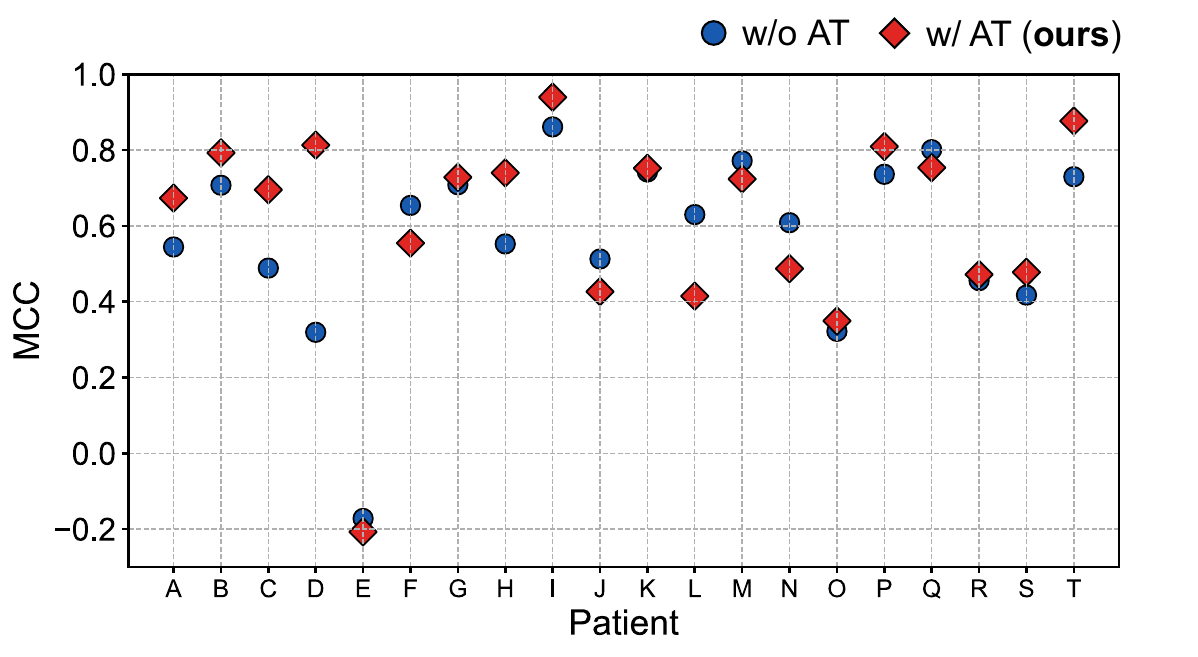} 
    \caption{Patient-specific MCC comparison between with and without adversarial training (AT)} 
    \label{fig:MCC} 
\end{figure}

\subsection{Baseline Comparisons}
To evaluate the effectiveness of temporal modeling, we compared two architectural variants: a baseline CNN and a CNN-BiLSTM model.
For each architecture, we assessed the impact of adversarial training by including or removing the first-stage domain classifier, while maintaining all other components and training procedures identical to the proposed method (CNN-BiLSTM with adversarial training).

\section{Results and Discussion}
Fig.~\ref{fig:Loss} shows the loss progression during the first stage of the proposed training process.
The results were computed using a specific random seed, with mean and standard deviation calculated across patients for each epoch.
The loss of the label predictor showed a consistent decrease throughout the training process. 
Conversely, the loss of the domain classifier exhibited a gradual increase over epochs,  indicating that the feature extractor successfully generated features that made domain discrimination increasingly challenging.
These results suggest that the feature extractor simultaneously accomplished two objectives: learning discriminative features to enhance seizure detection accuracy while achieving domain-invariance (i.e., patient-invariance) through adversarial training.
This demonstrates the effectiveness of adversarial training in multi-domain settings.

Table~\ref{AT_results} presents the impact of adversarial training on model performance, with bold values indicating the maximum scores for each metric. 
Overall, the CNN-BiLSTM model outperformed the CNN model.
While the CNN extracts features from EEG recordings at specific time points, relying solely on these features for seizure detection makes the model susceptible to noise, such as transient amplitude increases.
In contrast, the integration of CNN and BiLSTM enables the model to consider the temporal context by incorporating both past and future features in the sequence.
Given that seizures typically persist for tens of seconds, this temporal consideration effectively reduces false positives, leading to improved overall performance.

The application of adversarial training improved performance in both model types, with particularly notable effects in the proposed CNN-BiLSTM model. 
This improvement can be attributed to the ability of BiLSTM to capture temporal information, thereby reducing the impact of noise.
Consequently, the influence of individual differences became more prominent, highlighting the improvement in generalization capability achieved through adversarial training.
Fig.~\ref{fig:results} compares the seizure prediction probabilities and detection results obtained from the CNN-BiLSTM model with and without adversarial training. 
Fig.~\ref{fig:results}(a) shows an example where high seizure detection accuracy is achieved regardless of adversarial training, which is typical for patients with minimal domain shift.
In contrast, Fig.~\ref{fig:results}(b) presents a case where detection accuracy substantially decreases without adversarial training, demonstrating the effectiveness of the proposed method in extracting patient-invariant features.
This improvement can be attributed to the reduction in individual variations, which enhances the distinction between seizures and non-seizure states.

Fig.~\ref{fig:MCC} compares the MCC values obtained from the CNN-BiLSTM model for each patient with and without adversarial training.
Overall, the proposed method improved MCC for a majority of patients.
However, individual analysis revealed considerable variation across patients, with patients achieving higher MCC without adversarial training.
Without adversarial training, the model performance is strongly influenced by individual differences.
Consequently, the model tends to perform better when characteristics of the target patient coincidentally resemble those of source patients whose seizure and non-seizure states were accurately classified during training.
This observation explains why some patients showed higher MCC without adversarial training.
In contrast, adversarial training is designed to capture patient-invariant features, making the results less dependent on inter-patient similarities. 
By prioritizing features common across all patients over patient-specific features, this approach enables the development of a more generalizable model.  
This enhanced generalization capability appears to be the primary factor contributing to the overall performance improvement.

In summary, adversarial training effectively captured patient-invariant features, reduced the influence of individual differences, and improved overall performance. 
These results demonstrate the effectiveness of adversarial training in enhancing seizure detection capability. 
However, we observed persistently low MCC values for Patient E regardless of adversarial training.
This patient, at five months of age and the youngest in the dataset, exhibited EEG patterns that differed substantially from those of other patients.
This finding suggests that the current model and training approach have limitations in generalizing to EEG characteristics not represented in the training data, suggesting the need for further improvements in model robustness.

\section{Conclusion}
In this paper, we proposed a patient-invariant epileptic seizure detection method using adversarial training with a combined CNN-BiLSTM model.
The two-stage training process involves domain adversarial training in the first stage to extract patient-invariant features, followed by BiLSTM integration for temporal information processing.
Experimental results on 20 focal epilepsy patients demonstrated that our approach effectively reduces the impact of individual differences through adversarial training, while capturing temporal dependencies with BiLSTM.
However, challenges remain in accurately detecting seizures when EEG patterns of new patients differ significantly from others, particularly in early infancy patients.

In future work, we aim to develop more generalized and robust patient-invariant features while expanding the dataset to include more patients and longer EEG recordings. 
Additionally, we will optimize the BiLSTM time window to reduce detection delays and improve clinical applicability.

\addtolength{\textheight}{-12cm} 

\bibliographystyle{IEEEtran.bst}
\bibliography{EMBC2025.bib}

\end{document}